\documentclass{ws-mpla}
\begin{document}
\markboth{Abramovsky V.A.,Dmitriev A.V}{Multipomeron quasi-eikonal model for single diffraction processes}
\title{Multipomeron quasi-eikonal model for single diffraction processes}
\author{\footnotesize V.A.ABRAMOVSKY, A.V. DMITRIEV}
\address{Novgorod State University, B. S.-Peterburgskaya Street 41,\\
Novgorod the Great, Russia,\\ 173003}
\maketitle

\begin{abstract}
Single diffraction processes was usually treated in the triple-reggeon framework, but   this formalism is inconsistent with
CDF data. In this paper we show, that multipomeron quasi-eikonal model gives agreement with these data. Cross-section of
single diffraction processes at LHC energy is estimated.
\end{abstract}
\begin{figure}[th]
\centerline{\psfig{file=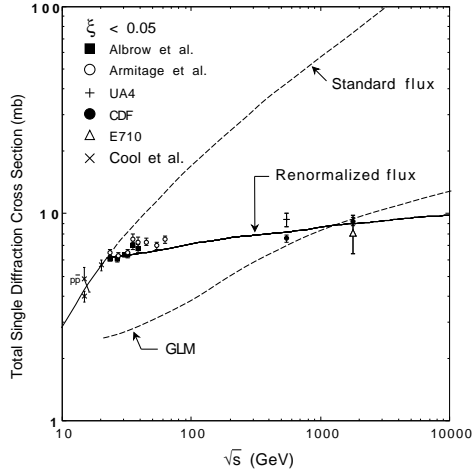,width=3in}}
\vspace*{0pt}
\caption{The total single diffraction cross section for $p(\bar{p})+p\rightarrow p(\bar{p})+X$ vs $\sqrt{s}$ compared with the predictions of the renormalized pomeron flux model of  Goulianos${}^3$ (solid line) and of the model of Gotsman, Levin and Maor${}^2$ (dashed line, labeled GLM); the latter, which includes "screening corrections", is normalized to the average value of the two CDF measurements at $\sqrt s=546$ and 1800 GeV.}
\label{fig:figgoul}
\end{figure}

Triple-reggeon phenomenology well describes total,elastic and low-energy diffraction data, but at the regeon of Tevatron
energies it fails to describe data on single diffraction dissociation. The main problem is that total single diffraction
cross-section rise more weaker than it is predicted by $Y$-like Regge diagrams, involved only three pomerons. This fact is
clearly seen from Fig.\ref{fig:figgoul} extracted from Ref.[\cite{goul_pic}], there "Standard flux" is corresponded to $Y$-like Regge diagram.

\begin{figure}[th]
\centerline{\psfig{file=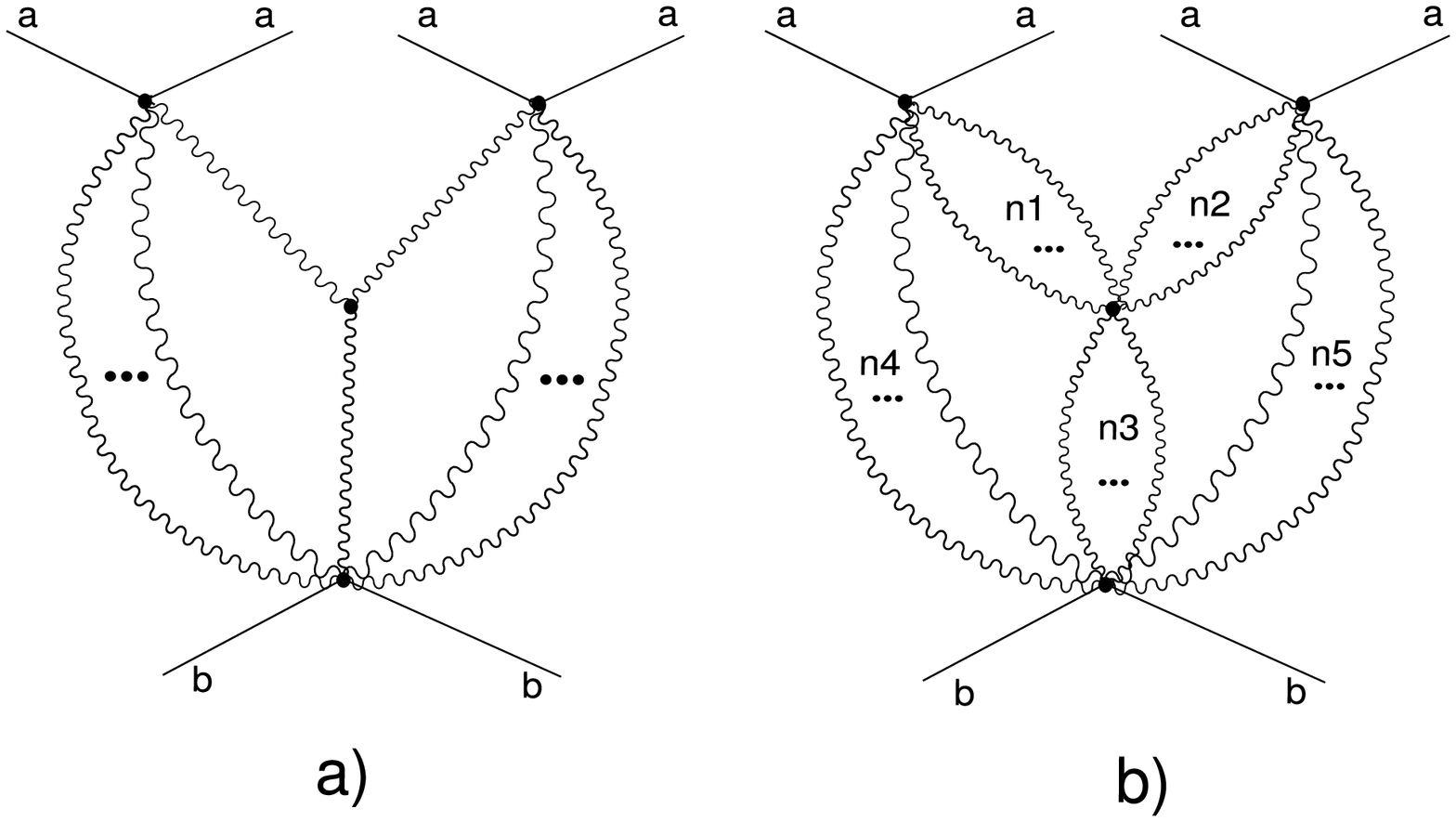,width=4in,angle=90}}
\vspace*{0pt}
\caption{Regge diagrams describing single diffraction dissociation of particle b.}
\label{fig:fig2}
\end{figure}

Many ways were suggested to solve this problem. First way is two-variant (Ref.[\refcite{goul_orig}]
and Ref.[\refcite{erhan_orig}]) pomeron flux renormalization model. This phenomenological approach well describes CDF
data on single diffraction, but we need more theoretical basises for extrapolation to higher energies.
Second way is straight-forward account of screening corrections (Ref.[\refcite{gotsman_orig}] and Ref.[\refcite{chung_orig}]).
This way seems to be more natural, but we need to introduce additional parameters and make some assumptions about Regge
diagram technics. In Ref.[\refcite{gotsman_orig}] and Ref.[\refcite{chung_orig}] only some part of sufficient diagrams  was
accounted, see Fig.\ref{fig:fig2}a, but central $Y$-like diagram was modified to account processes at low-energy regeon.

Because low-energy corrections rapidly decrease with energy, we account only pomeron contributions, but in all
sufficient diagrams, as it was done in Ref.[\refcite{all}] and Ref.[\refcite{Ab}]. It gives us possibility to normalize cross-section of the single diffraction to CDF data and
make theoreticaly based predictions for cross-section of the single diffraction at LHC energies.

In this paper we account all non-enhanced absorptive corrections to the Y-diagram
contribution. shown in  Fig.\ref{fig:fig2}b. We use quasi-eikonal approach, where the vertex of the interaction of n
pomerons with hadron is written as
\begin{equation}
N_h(k_1,..,k_n)=g_h(g_hc_h)^{n-1}exp\left(-R_h^2\sum_{i=1}^{n}k_i^2\right)
\end{equation}
and the vertex corresponding to the transition of l pomerons into m pomerons under the $\pi$ -meson exchange dominance assumption is
\begin{equation}
\Lambda(k_1,..,k_{l+m})=r(g_{\pi}c_{\pi})^{l+m-3}exp\left(-R_r^2\sum_{i=1}^{l+m}k_i^2\right).
\end{equation}
Here $g_h$ is the pomeron-hadron coupling, $c_h$ is the corresponding shower enhancement coefficient, $R_h$ and $R_r$ are
the radii of the pomeron-hadron and pomeron-pomeron interactions, respectively, $k_i$ are the pomeron transverse momenta.
The contribution $f_{n_1 n_2 n_3 n_4 n_5}$ of each diagram in Fig.\ref{fig:fig2}b can be written in the rather simple form
\begin{equation}
\begin{array}{l}
f_{n_1 n_2 n_3 n_4 n_5}=\frac{(-1)^{n_1+n_2+n_3+n_4+n_5+1}}{n_1!n_2!n_3!n_4!n_5!}\frac{8 \pi^3 r}{c_a^2 c_b g_{\pi} c_{\pi}}
\left[\frac{g_a c_a g_{\pi} c_{\pi} e^{\Delta (Y-y)}}
{8\pi (R_a^2+R_{\pi}^2+\alpha^{\prime}(Y-y))}\right]^{n1+n2} \\
\\
\left[\frac{g_a c_a g_b c_b e^{\Delta Y}}
{8\pi (R_a^2+R_b^2+\alpha^{\prime}Y)}\right]^{n4+n5}
\left[\frac{g_b c_b g_{\pi} c_{\pi} e^{\Delta y}}
{8\pi (R_b^2+R_{\pi}^2+\alpha^{\prime}y)}\right]^{n3}
\frac{1}{detF}e^{-t\frac{c}{detF}} \\
\\
detF=a_1 a_2 a_3 + a_1 a_3 a_5 + a_1 a_2 a_5 + a_1 a_2 a_4 + a_2 a_3 a_4 + a_1 a_4 a_5 + a_3 a_4 a_5 + a_2 a_4 a_5 \\
c=a_2 a_3 + a_1 a_5 + a_3 a_5 + a_2 a_5 + a_1 a_3 + a_1 a_4 + a_3 a_4 + a_2 a_4 \\
a_1=\frac{n_1}{R_a^2+R_{\pi}^2+\alpha^{\prime}(Y-y)}\\
a_2=\frac{n_2}{R_a^2+R_{\pi}^2+\alpha^{\prime}(Y-y)}\\
a_3=\frac{n_3}{R_b^2+R_{\pi}^2+\alpha^{\prime}y}\\
a_4=\frac{n_4}{R_a^2+R_b^2+\alpha^{\prime}Y}\\
a_5=\frac{n_5}{R_a^2+R_b^2+\alpha^{\prime}Y}.
\end{array}
\end{equation}
Here $Y=ln(s)$ $y=ln(M^2)$.
Then the inclusive cross section may be written as
\begin{equation}
(2\pi)^3 2E \frac{d^3\sigma }{dp^3 }=\pi \frac{s}{M^2}\sum_{n_1,n_2,n_3=1}^{\infty} \sum_{n_4,n_5=0}^{\infty}f_{n_1 n_2 n_3 n_4 n_5}
\end{equation}

To estimate  inclusive cross section we use parameters describing the data on the total and elastic cross section:
$\Delta=0.12$ $g_p^2=60.3 GeV^2$ $\alpha^{\prime}=0.25 GeV^{-2}$ $c_p^2=1.6$ $c_{pi}^2=1.6$ $R_p^2=1.78 GeV^2$
$R_r^2=0.62 GeV^2$ $g_{\pi}=2/3g_p$. Parameter $r$ is freely varied, because total and elastic cross section does not
depend on it.

\begin{figure}[th]
\centerline{\psfig{file=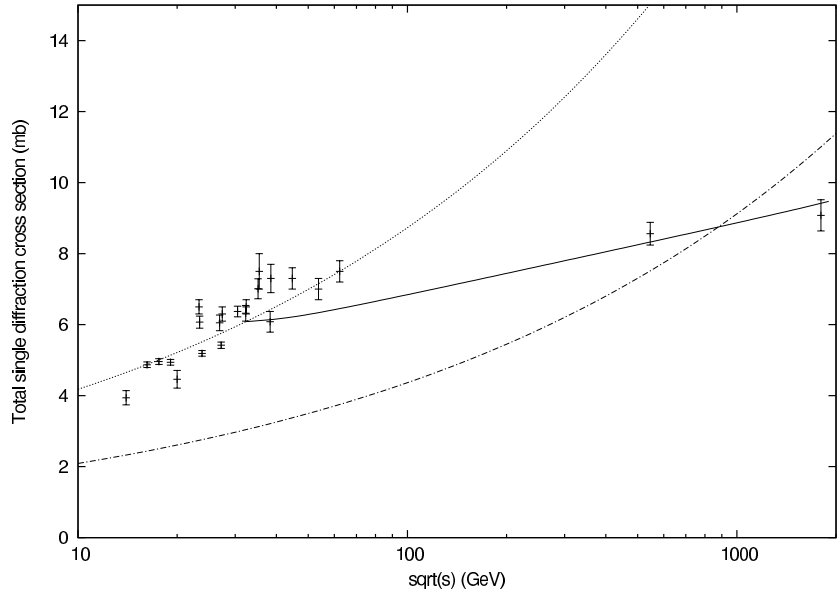,width=5in}}
\vspace*{8pt}
\caption{Triple-reggeon (dashed and dot-dashed) predictions vs eikonaliezed (solid line) predictions for energy dependence
 of total cross-section of single diffraction.}
\label{fig:fig1}
\end{figure}

Obtained dependence of $\sigma_{SD}$ is shown on Fig.\ref{fig:fig1} as solid line with compilation of experimental data set taken from Ref.[\refcite{goul_pic}]. CDF and UA4 points at $\sqrt{s}=546 GeV$ and CDF and E710 points at $\sqrt{s}=1800 GeV$ was united.

High-energy CDF data is well described by our model and low-energy data don`t contradict to our estimations, if we
 assume, that low-lying reggeons was neglected. Comparing Fig.\ref{fig:figgoul} and \ref{fig:fig1} we see, that our
  predictions is similar to Goulianos predictions. We can examine our model as Regge based explanation pomeron-flux
   renormalization factor, which was entered as axiom by Goulianos \cite{goul_orig}. This factor, defined as ratio
    $k_H(s,M^2,t)$ of sum of all diagrams with absorptive corrections to $Y$-like diagram was calculated in
     Ref.[\refcite{Ab}].  It was shown \cite{Ab}, that $k_H(s,M^2,t)$ slowly varying with $M^2$ and $t$ and in this work
      we show, that $k_H(s)$ correctly describe existed data. In the terms of pomeron-flux renormalization models our
       calculations makes Goulianos model \cite{goul_orig} more attractive, than Erhan-Shlein one \cite{erhan_orig}.

Our predictions for higher energies region notably differs from other models.
At LHC the centre of mass energy for protons will be 14TeV. We predict, that at this energy integrated single diffraction
cross-section will be about
\begin{equation}
\sigma_{SD}(\sqrt{s}=28 TeV)=12.6 mb
\end{equation}
Goulianos model \cite{goul_orig} predicts at this energy $\sigma_{SD}(\sqrt{s}=28 TeV)=10 \pm 0.5 mb$.
Model of Gotsman, Levin and Maor predicts \cite{gotsman_orig} $\sigma_{SD}(\sqrt{s}=28 TeV)=13.3 \pm 0.3 mb$.

\section*{Acknowledgments}
We thank N.Prikhod`ko for useful discussions.
This work was supported by RFBR Grant RFBR-03-02-16157a and grant of Ministry for Education E02-3.1-282

\end{document}